\documentclass[12pt,a4paper]{article}
\usepackage[latin1]{inputenc}
\usepackage{amsmath}
\usepackage{braket}
\usepackage{amsfonts}
\usepackage{amssymb}
\usepackage{graphicx}
\usepackage{amsthm}
\usepackage{listings}
\usepackage{float}
\usepackage{caption}
\usepackage{color}
\usepackage{chngcntr}
\usepackage{cite}
\usepackage{authblk}
\usepackage{bbold}
\usepackage{braket}   
\usepackage{subfigure}
\usepackage[english]{babel}  
\usepackage[left=2.00cm, right=2.00cm, top=2.00cm, bottom=2.00cm]{geometry}

\title{Dirac electron in graphene with magnetic fields arising from first-order intertwining operators}
\date{ }
\author{M Castillo-Celeita\footnote{mfcastillo@fis.cinvestav.mx} }
\author{David J. Fern\'andez C.\footnote{david@fis.cinvestav.mx}}
\affil{\small Physics Department, Cinvestav, P.O. Box. 14-740, 07000 Mexico City, Mexico}

\begin{document}
\maketitle

\begin{abstract}
The behaviour of a Dirac electron in graphene, under magnetic fields which are orthogonal to the layer, is studied. The initial problem is reduced to an equivalent one, where two one-dimensional Schr\"{o}dinger Hamiltonians $H^{\pm}$ are intertwined by a first order differential operator. Special magnetic field are initially chosen, in order that $V^{\pm}$ will be shape invariant exactly solvable potentials. When looking for more general first order operators, intertwining $H^-$ with a non necessarily shape invariant Hamiltonian, new magnetic fields associated also to analytic solutions will be generated. The iteration of this procedure is as well discussed.          
 
\end{abstract}

\section{Introduction}
Graphene is a two dimensional material composed of a single layer of carbon atoms placed in a hexagonal arrangement, which has remarkable mechanical and electrical properties: its mechanical resistance is greater than for the steel, it is flexible, it has high electrical conductivity and it is transparent, among other characteristics \cite{nbot17}. Despite graphene was analyzed theoretically in the fifth decade of the previous century~\cite{w47}, this description remained unnoticed until this material was isolated at room temperature by Geim and Novoselov in 2004~\cite{NGMJZ, k07, ngp09}. Since then the scientific comunity has improved the graphene production by ever larger quantities with the use of different methods~\cite{h08, pc12}. On the theoretical side, graphene is important in different fields such as  solid state physics, quantum mechanics and field theory, because it can model a massless Dirac electron at low velocity ($\sim10^{-2}c$) and energy~\cite{NG, NGMJZ2}. This led to proposals for testing interesting phenomena in graphene, such as the Klein tunneling or the quantum Hall-Effect~\cite{KNG, JNP, zts05, wdm07}, where external magnetic fields are necessarily involved.\\

It is known nowadays that magnetic fields are a good option to confine electrons in a space region~\cite{wdm07, mde09, b06, mrsr10, rr10}. In fact, the interaction of the graphene's massless Dirac electron with external magnetic fields is modeled theoretically by the Dirac-Weyl equation (DWE), and the same happens for other carbon allotropes as the carbon nanotubes and fullerenes \cite{df17,jknt13,jkn14,knt15}. This problem has been solved exactly for graphene in some inhomogeneous magnetic fields~\cite{em17}. Moreover, since the DWE can be reduced to a pair of stationary Schr\"{o}dinger equations intertwined by first-order differential operators, then supersymmetric quantum mechanics (SUSY QM) is a natural tool to find exact solutions for the problem. In this method it is customary to generate a potential with the same spectrum as the initial one, except by the ground state energy, for the so-called shape invariant potentials~\cite{knn09,mtm11,chrv18,jss18,jk14, j15}. A generalization of this method was made by Midya and Fernandez~\cite{mf14}, where the initial magnetic field was deformed by introducing one parameter $\lambda$, so that the new potentials are no longer shape invariant unless $\lambda\rightarrow \infty$. Besides, they added another parameter $\delta$ to the new Hamiltonian, so that its energy levels are moved with respect to the initial ones when $\delta$ is changed.\\       

In this work we will extend the treatment introduced in~\cite{mf14}, making use of solutions of the initial Schr\"{o}dinger equation instead of solutions of the corresponding Riccati equation to generate the new potential; by approaching in this way the problem it will be clear how to iterate the method, and we will call these first order potentials. Later on we will generate the second order potentials by iterating the method. At the end, it will be shown how to generate the kth order potentials.

The paper is organized as follows: in section 2 we will discuss how to implement the SUSY QM to solve the DWE. In section 3 the initial potentials will be deformed through the techniques mentioned above; in the same section the method will be extended to second order and also generalized to order $k$. In section 4 we will apply the technique to two particular examples: the harmonic oscillator and the Morse potential. The last section contains our conclusions.       

\section{DWE and SUSY QM}\label{sec2}

The DWE describes the behaviour of a massless Dirac electron in graphene at low energies. When applying a magnetic field $\bf B$ this equation appears from the free case with minimal coupling:
    
\begin{equation}
v_F\mathbf{\sigma}.\bigg(\mathbf{p}+\frac{e\mathbf{A}}{c}\bigg)\Psi(x,y)=E\Psi(x,y),
\label{c2eq1}
\end{equation}

\noindent
where $v_F=c/300$ is the Fermi velocity, $\mathbf{\sigma}=(\sigma_x, \sigma_y)$ are the Pauli matrices,  $\mathbf{p}=-i\hbar(\partial_x,\partial_y)$ is the momentum operator in coordinates representation, and the vector potential in the Landau Gauge will be taken as $\mathbf{A}=(0,A_y(x),0)$, which produces a magnetic field $\mathbf{B}=\nabla\times\mathbf{A}$ pointing along $z$ direction. The electron motion takes place on the $x$-$y$ plane, which is characterized by an associated spinor wave function $\Psi(x,y)$. Besides, the previous choices imply that in the $y$ axis a free motion appears, which entails the separation of variables:

\begin{equation}
\Psi(x,y)= e^{iky}\bigg[\begin{array}{c}
\psi^+(x)  \\ 
i\psi^-(x)
\end{array} \bigg],
\label{c2eq2}
\end{equation} 
 
\noindent
with $k$ being the wavenumber in $y$ direction. Equations~(\ref{c2eq1}) and~(\ref{c2eq2}) lead to a pair of coupled differential equations:

\begin{equation} 
\begin{bmatrix}
0& \frac{d}{dx}+(k+\frac{eA_y}{c\hbar})\\
-\frac{d}{dx}+(k+\frac{eA_y}{c\hbar})&0\\
\end{bmatrix}
\begin{bmatrix}
\psi^+(x)  \\ 
i\psi^-(x) 
\end{bmatrix}
=\frac{E}{\hbar v_F}
\begin{bmatrix}
\psi^+(x)\\ 
i\psi^-(x)
\end{bmatrix}
\end{equation}

\noindent
which, when decoupled, produce two independent Schr\"{o}dinger equations:

\begin{equation}
H^{\pm}\psi^{\pm}(x,y)=\bigg(-\frac{d^2}{dx^2}+V^{\pm}(x)\bigg)\psi^{\pm}(x)={\cal{E}}\psi^{\pm}(x),
\label{c2eq3}
\end{equation}

\noindent  
where the potentials $V^{\pm}(x)$ and energy ${\cal{E}}$ are given by:

\begin{equation}
V^{\pm}(x)=\bigg(k+\frac{eA_y(x)}{c\hbar}\bigg)^2\pm\frac{e}{c\hbar}\frac{dA_y(x)}{dx},~~~~{\cal{E}}=\frac{E^2}{v^2_F\hbar^2}.
\label{c2eq4}
\end{equation}

\noindent  
The form of these equations suggests to use SUSY QM to deal with the problem \cite{cks95,ba00,cr00,acin00,fe10,fe19}. 

\subsection{SUSY QM and shape invariant potentials} 

The Hamiltonians in Eq.~(\ref{c2eq3}) are factorized as follows:
 
\begin{equation}
H^{\pm}=L_0^{\mp}L_0^{\pm},
\end{equation}

\noindent
where the intertwining operators $L_0^{\mp}$, which satisfy the relations $H^{\pm}L_0^{\mp}=L_0^{\mp}H^{\mp}$, are given by:

\begin{equation}
L_0^{\pm}=\mp\frac{d}{dx}+W_0(x),
\end{equation}

\begin{equation}
W_0(x)=\frac{eA_y(x)}{c\hbar}+k,
\end{equation}

\noindent
with $W_0(x)$ being the superpotential, which allows to express the potentials (\ref{c2eq4}) in compact form as follows:
\begin{equation} 
V^{\pm}(x)=W_0^2(x)\pm W_0'(x).
\label{c2eq5}
\end{equation}
The action of the intertwining operators $L_0^{\pm}$ on the eigenstates of the Hamiltonians~(\ref{c2eq3}) become:
\begin{equation} 
\psi^+_{n}(x)=\frac{L_0^-\psi^-_{n+1}(x)}{\sqrt{{\cal{E}}_{n+1}^-}},~~~\psi^-_{n+1}(x)=\frac{L_0^+\psi^+_n(x)}{\sqrt{{\cal{E}}_n^+}},~~~\psi_0^-(x)\sim e^{-\int W_0(x)dx},
\end{equation}

\noindent
where the operator $L_0^-$ annihilates the ground state of $H^-$, $L_0^-\psi_0^-(x)=0$, which implies that $W_0(x)=-{\psi^-_0(x)}'/\psi^-_0(x)$. The energy levels of $H^{\pm}$ turn out to be: 

\begin{equation}
{\cal{E}}_n^+={\cal{E}}_{n+1}^-,~~~{\cal{E}}_0^-=0.
\end{equation}

\noindent
These expressions indicate that the eigenfunctions and eigenvalues of the problem can be found through the operators $L_0^{\pm}$, which simplifies the calculations since this involves just first-order derivatives.\\
Note that the magnetic field amplitude $B_0(x)$ is connected with the $y$ component of the vector potential, with the superpotential $W_0(x)$ and with the ground state of $H^-$ as follows:

\begin{equation}
B_0(x)=\frac{dA_y(x)}{dx}=\frac{c\hbar}{e}\frac{dW_0(x)}{dx}=-\frac{c\hbar}{e}\frac{d^2}{dx^2}\{\ln[\psi_0^-(x)]\}.
\end{equation} 

\noindent
In addition, for some special forms of $B_0(x)$ the two SUSY partner potentials $V^{\pm}(x)$ become shape invariant \cite{knn09,mtm11,chrv18,jss18,jk14,j15,cks95,ba00,cr00}, which implies that $H^{\pm}$ are exactly solvable so that the eigenfunctions and eigenvalues are known explicitly.  
Therefore, from a shape invariance point of view the problem is essentially solved in these cases. However, it is still feasible to generalize the method (see also \cite{mf14,fh99}), and to iterate it as many times as we wish, which will be shown in the following sections.   

\section{Generalized intertwining and iterations}\label{sec3}
\subsection{Generalized first order intertwining}

The first step of this method is to displace up the energy of the Hamiltonian $H^-$ as follows,

\begin{equation}
\tilde{H_0}\equiv H^--\epsilon_1=-\frac{d^2}{dx^2}+V^-(x)-\epsilon_1,
\label{c3eq0}
\end{equation}

\noindent
so that $\tilde{V}_0(x)\equiv V^-(x)-\epsilon_1$, where $\epsilon_1\leq{\cal{E}}_0^-=0$. The second step is to build a new Hamiltonian $H_1$ departing from $\tilde{H}_0$ through the intertwining relation:

\begin{equation}
H_1L^+_1=L_1^+\tilde{H}_0,
\label{c3eq1}
\end{equation}

\noindent
where $H_1$ and $L_1^{\pm}$ are given by:

\begin{equation}
H_1=-\frac{d^2}{dx^2}+V_1(x,\epsilon_1),~~~~L_1^{\pm}=\mp\frac{d}{dx}+W_1(x,\epsilon_1).
\end{equation}

\noindent
Equation~(\ref{c3eq1}) implies that the Hamiltonians $\tilde{H}_0$ and $H_1$ are factorized as $\tilde{H}_0=L_1^-L_1^+$ and $H_1=L_1^+L_1^-$, which leads to the following Riccati equation and relation between potentials \cite{ff05}:

\begin{equation}
W_1^2(x,\epsilon_1)+W'_1(x,\epsilon_1)=\tilde{V}_0(x),
\label{c3eq2}
\end{equation}

\begin{equation}
V_1(x,\epsilon_1)=\tilde{V}_0(x)-2 W_1'(x,\epsilon_1).
\label{c3eq3}
\end{equation}

\noindent
Let us suppose now that $W_1(x,\epsilon_1)=u_1^{(0)'}/u_1^{(0)}$; thus Eq.~(\ref{c3eq2}) is transformed into:
 
\begin{equation}
-u_1^{(0)''}+\tilde{V}_0(x)u_1^{(0)}=0,
\label{c3eq10}
\end{equation}

\noindent
which is the stationary Schr\"{o}dinger equation associated to $\tilde{H}_0$ for null factorization energy. A nodeless solution to this equation constitutes the information required to build the potential $V_1(x,\epsilon_1)$ \cite{ff05} (see also equation~(\ref{c3eq3})). Moreover, the magnetic field giving place to $V_1(x,\epsilon_1)$ is calculated as follows:

\begin{equation}
B_1(x,\epsilon_1)=\frac{c\hbar}{e}\frac{dW_1(x,\epsilon_1)}{dx}=-B_0(x)+\frac{c\hbar}{e}\frac{d^2}{dx^2}\{\ln[u_1^{(0)}/\psi_0^-(x)]\}.
\end{equation}

The third step of our method is to identify the eigenfunctions and eigenvalues of the new system. The energy levels for $\tilde{H}_0$ and $H_1$ are those of $H^-$, displaced by the quantity $-\epsilon_1$, plus the ground state of $H_1$ at zero energy:

\begin{equation}
\begin{split} 
&\tilde{\cal{E}}_n^{(0)}={\cal{E}}_n^--\epsilon_1, \\
&{\cal{E}}_0^{(1)}=0,~~~{\cal{E}}_{n+1}^{(1)}=\tilde{\cal{E}}_n^{(0)},~~n=0,1,\dots
\end{split} 
\end{equation}

\noindent
with $\epsilon_1\leq{\cal{E}}_0^-=0$. A diagram of these energies can be seen in figure \ref{fig1}. The unknown eigenfunctions associated to the previous energies are given by: 

\begin{equation}
\psi_0^{(1)}(x)\sim e^{-\int W_1(x,\epsilon_1)dx}=\frac{1}{u_1^{(0)}},~~~~\psi_{n+1}^{(1)}(x)=\frac{1}{\sqrt{\tilde{\cal{E}}_n^{(0)}}}L_1^+\psi_n^-(x),
\end{equation}

\noindent
where the eigenfunctions $\psi_n^-(x)$ of $H^-$, and consequently those of $\tilde{H}_0$, are assumed to be known. In addition, the ground state of $H_1$ fulfills the condition $L_1^-\psi_0^{(1)}(x)=0$.\\

Note that a similar generalization was proposed in \cite{mf14}, where $H^+$ was displaced as in Eq.(\ref{c3eq0}) instead of $H^-$ and it was also supposed that $\epsilon_1\equiv-\delta\leq{\cal E}_0^-\Rightarrow\delta\geq 0$. However, since the non-singular first-order SUSY transformations can be implemented for factorization energies below the ground state energy of the initial Hamiltonian, for $\epsilon_1\leq{\cal E}_0^+$ in the case addressed in \cite{mf14}, then the non-singular domain $\delta\geq -{\cal E}^+_0$ is really larger that the one reported there, which should include as well the interval $\delta\in[-{\cal E}_0^+,0)$. Moreover, the search of the general solution of the key equations (\ref{c3eq2},\ref{c3eq10}) of this treatment looks simpler for the Schr\"{o}dinger equation (as in this paper) than for the Riccati equation (as in \cite{mf14}), although both are indeed equivalent (remember that we can go from the Riccati equation (\ref{c3eq2}) to the Schr\"{o}dinger equation (\ref{c3eq10}) through the change $W_1=[\log u_1^{(0)}]'$, and vice versa by using $u_1^{(0)}\propto \exp[\int W_1 dx]$). In addition, we will see next that our approach allows to guess simply how to iterate the method in a straightforward way, which was not seen in \cite{mf14}.       

\subsection{Iteration of the method: second intertwining}

\begin{figure}
\centering
\includegraphics[scale=1.0]{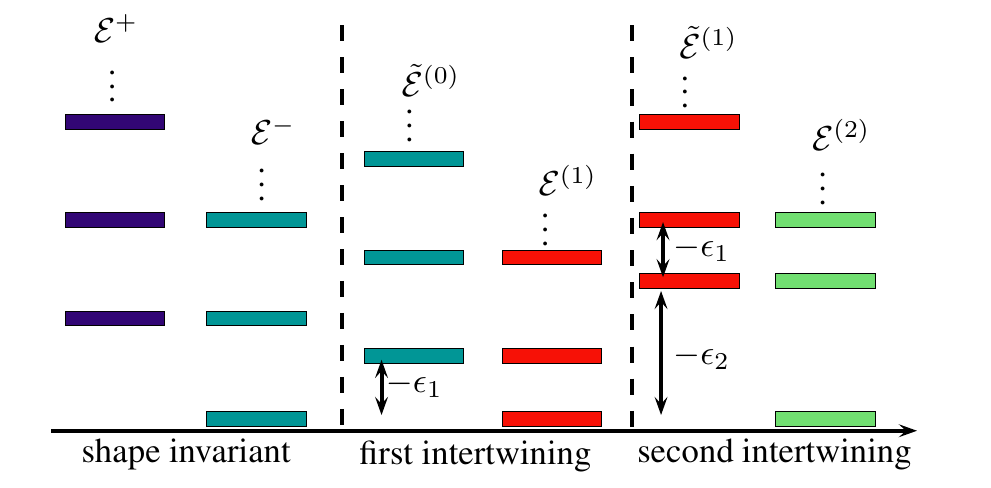}
\caption{Representation of the energy levels for the chain of Hamiltonians.}
\label{fig1}
\end{figure}

Let us repeat now the steps pointed out at the previous section: in the first place a factorization energy $\epsilon_2$ is subtracted to the Hamiltonian $H_1$: 

\begin{equation}
\tilde{H}_1\equiv H_1-\epsilon_2,
\end{equation}

\noindent
such that $\epsilon_2<\epsilon_1$. Thus, the ground state energy of $\tilde{H}_1$ is $\tilde{\cal{E}}_0^{(1)}=-\epsilon_2\geq 0$ (see figure \ref{fig1}), and we denote $\tilde{V}_1(x,\epsilon_1)\equiv V_1(x,\epsilon_1)-\epsilon_2$. The intertwining relation reads:

\begin{equation}
H_2L_2^+=L_2^+\tilde{H}_1,
\end{equation}

\noindent
with the new Hamiltonian $H_2$ and intertwining operators $L_2^{\pm}$ being given by:

\begin{equation}
H_2=-\frac{d^2}{dx^2}+V_2(x,\epsilon_2),~~~L_2^{\pm}=\mp\frac{d}{dx}+W_2(x,\epsilon_2).
\end{equation}

\noindent
The Hamiltonians $\tilde{H}_1$, $H_2$ are expressed in terms of the intertwining operators as $\tilde{H}_1=L_2^-L_2^+$ and $H_2=L_2^+L_2^-$, which leads to a pair of equations similar to~(\ref{c3eq2},~\ref{c3eq3}):

\begin{equation}
W_2^2(x,\epsilon_2)+W'_2(x,\epsilon_2)=\tilde{V}_1(x,\epsilon_1),
\label{c3eq4}
\end{equation}

\begin{equation}
V_2(x,\epsilon_2)=\tilde{V}_1(x,\epsilon_1)-2 W_2'(x,\epsilon_2).
\label{c3eq5}
\end{equation}

\noindent
Once again we pass from the Riccati equation~(\ref{c3eq4}) to the associated Schr\"{o}dinger equation through the change:

\begin{equation}
W_2(x,\epsilon_2)=\frac{u_2^{(1)'}}{u_2^{(1)}},
\label{c3eq6}
\end{equation}

\noindent
leading to: 

\begin{equation}
-u_2^{(1)''}+\tilde{V}_1(x,\epsilon_1)u_2^{(1)}=0.
\end{equation}

\noindent
The intertwining operator $L_1^+$ helps to find $u_2^{(1)}$, by connecting it with the corresponding Schr\"{o}dinger solution $u_2^{(0)}$ for $\tilde{H}_0$:

\begin{equation}
u_2^{(1)}\propto L_1^+u_2^{(0)}=-u_2^{(0)'}+\frac{u_1^{(0)'}}{u_1^{(0)}}u_2^{(0)}=-\frac{\mathsf{W}[u_1^{(0)},u_2^{(0)}]}{u_1^{(0)}},
\label{c3eq7}
\end{equation}
  
\noindent
where $\mathsf{W}[u_1^{(0)},u_2^{(0)}]$ denotes the Wronskian of $u_1^{(0)}$ and $u_2^{(0)}$ while $u_2^{(0)}$ fulfills:

\begin{equation}
-u_2^{(0)''}+V^-(x)u_2^{(0)}=(\epsilon_1+\epsilon_2)u_2^{(0)}.
\label{c3eq11}
\end{equation}
 
\noindent
By combining Eqs.~(\ref{c3eq5}) and~(\ref{c3eq7}) it is obtained:

\begin{equation}
\begin{split} 
&V_2(x,\epsilon_2)=\tilde{V}_0(x)-2\frac{d^2}{dx^2}\ln \mathsf{W}[u_1^{(0)},u_2^{(0)}]-\epsilon_2=V^-(x)-2\frac{d^2}{dx^2}\ln\mathsf{W}[u_1^{(0)},u_2^{(0)}]-(\epsilon_1+\epsilon_2).
\end{split} 
\end{equation}

\noindent
By means of Eqs.~(\ref{c3eq6}) and~(\ref{c3eq7}) the second-order magnetic field now can be found:

\begin{equation}
B_2(x,\epsilon_2)=\frac{c\hbar}{e}\frac{dW_2(x,\epsilon_2)}{dx}=-B_1(x,\epsilon_1)+\frac{c\hbar}{e}\frac{d^2}{dx^2}\{\ln[\mathsf{W}[u_1^{(0)},u_2^{(0)}]]\}.
\end{equation}

\noindent
The eigenvalues of $\tilde{H}_1$ and $H_2$ are related through:
\begin{equation}
\begin{split} 
&\tilde{\cal{E}}_n^{(1)}={\cal{E}}_n^{(1)}-\epsilon_2,\\
&{\cal{E}}_0^{(2)}=0,~~~{\cal{E}}_{n+1}^{(2)}=\tilde{\cal{E}}_n^{(1)}.
\end{split} 
\end{equation}

\noindent
The unknown eigenfunctions, associated with the last eigenvalues, are:

\begin{equation}
\psi_0^{(2)}(x)\sim e^{-\int W_2(x,\epsilon_2)dx}=\frac{1}{u_2^{(1)}},~~~\psi^{(2)}_{1}(x)=\frac{L_2^+\psi^{(1)}_{0}(x)}{\sqrt{\tilde{\cal{E}}_{0}^{(1)}}},~~~\psi_{n+2}^{(2)}(x)=\frac{L_2^+\psi^{(1)}_{n+1}(x)}{\sqrt{\tilde{\cal{E}}_{n+1}^{(1)}}}=\frac{L_2^+L_1^+\psi_n^-(x)}{\sqrt{\tilde{\cal{E}}_{n+1}^{(1)}\tilde{\cal{E}}_n^{(0)}}},
\end{equation}

\noindent
where $n=0,1,2\dots$

\subsection{$k$th intertwining}

For the $k$th intertwining the same scheme is followed: the Hamiltonian $H_{k-1}$ is first displaced by an amount $-\epsilon_k$:

\begin{equation}
\tilde{H}_{k-1}=H_{k-1}-\epsilon_k,
\end{equation}

\noindent
where $\epsilon_k<\epsilon_{k-1}<\dots\epsilon_2<\epsilon_1\leq 0$. The intertwining relation for the $k$th step is:

\begin{equation}
H_kL_k^+=L_k^+\tilde{H}_{k-1},
\label{c3eq8}
\end{equation}

\noindent
while the Hamiltonians $H_k$, $\tilde{H}_{k-1}$ and intertwining operator $L^{\pm}_k$ are expressed as:

\begin{equation}
\begin{split}
&H_k=-\frac{d^2}{dx^2}+V_k,~~~~\tilde{H}_{k-1}=-\frac{d^2}{dx^2}+V_{k-1}-\epsilon_k,\\
&L_k^{\pm}=\mp\frac{d}{dx}+W_k(x,\epsilon_k).
\label{c3eq9}
\end{split}
\end{equation}

\noindent
The factorizations involved are now $\tilde{H}_{k-1}=L_k^-L_k^+$ and $H_k=L_k^+L_k^-$, while the equations resulting from~(\ref{c3eq8}) and~(\ref{c3eq9}) become:

\begin{equation}
W^2_k(x,\epsilon_k)+W'_k(x,\epsilon_k)=\tilde{V}_{k-1},
\end{equation}

\begin{equation}
V_k=\tilde{V}_{k-1}-2W_k'(x,\epsilon_k).
\end{equation}

\noindent
If the superpotential is expressed as $W_k(x,\epsilon_k)=u^{(k-1)'}_k/u^{(k-1)}_k$, then it is obtained the following Schr\"{o}dinger equation for zero factorization energy:

\begin{equation}
-u^{(k-1)''}_k+\tilde{V}_{k-1}u^{(k-1)}_k=0.
\end{equation}

\noindent
The intertwining operators supply the relation  between $u^{(k-1)}_k$ and its predecessor, i.e., $u^{(k-1)}_k\propto L^+_{k-1}u^{(k-2)}_k$, where $u^{(k-2)}_k$ fulfills the Schr\"{o}dinger equation:  

\begin{equation}
-u_k^{(k-2)''}+V_{k-2}u_k^{(k-2)}=(\epsilon_k+\epsilon_{k-1})u_k^{(k-2)}.
\end{equation}

\noindent
By iterating these expressions for lower indices it turns out that $u^{(k-1)}_k\propto L^+_{k-1}L^+_{k-2}...L^+_1 u^{(0)}_k$, where

\begin{equation}
-u_k^{(0)''}+V^-(x)u_k^{(0)}=\bigg(\sum_{i=1}^k\epsilon_i\bigg) u_k^{(0)},
\end{equation}

\noindent
which allows to know all the superpotentials as well as the new potential:

\begin{equation}
V_k=V^-(x)-2\frac{d^2}{dx^2}\{\ln[\mathsf{W}(u_1^{(0)},\dots,u_k^{(0)})]\}-\bigg(\sum_{i=1}^k\epsilon_i\bigg).
\end{equation}

\noindent
The $kth$-order magnetic field reads now:

\begin{equation}
B_k(x,\epsilon_k)=\frac{c\hbar}{e}\frac{d}{dx}W_k(x,\epsilon_k)=-B_{k-1}(x,\epsilon_{k-1})+\frac{c\hbar}{e}\frac{d^2}{dx^2}\{\ln[\mathsf{W}(u^{(k-2)}_{k-1},u^{(k-2)}_k)]\}.
\end{equation}

\noindent
A useful form of the previous expression is also given by:

\begin{equation}
B_k(x,\epsilon_k)=\frac{c\hbar}{e}\frac{d^2}{dx^2}\bigg\{\ln\bigg[\frac{\mathsf{W}(u_1^{(0)},\dots,u^{(0)}_k)}{\mathsf{W}(u_1^{(0)},\dots,u^{(0)}_{k-1})}\bigg]\bigg\}.
\end{equation}

As previously, the eigenvalues and eigenfunctions of $H_k$, $\tilde{H}_{k-1}$ must be identified. The general relation between eigenvalues is:

\begin{equation}
\begin{split} 
&\tilde{\cal{E}}_{n}^{(k-1)}={\cal{E}}_n^{(k-1)}-\epsilon_k,\\
&{\cal{E}}_0^{(k)}=0,~~~~{\cal{E}}_{n+1}^{(k)}=\tilde{\cal{E}}_n^{(k-1)}.
\end{split} 
\end{equation}

\noindent
The eigenfunctions associated to the last eigenvalues are given by:
\begin{equation}
\begin{split}
&\psi_0^{(k)}(x)\sim e^{-\int W_k(x,\epsilon_k)dx}, \\
&\psi_1^{(k)}(x)=\textstyle{\frac{1}{\sqrt{{\cal{E}}_0^{(k-1)}-\epsilon_k}}}L_k^+\psi^{(k-1)}_0(x),\\
&\vdots \\
&\psi_{k-1}^{(k)}(x)=\textstyle{\frac{1}{\sqrt{({\cal{E}}_{k-2}^{(k-1)}-\epsilon_k)...({\cal{E}}_0^{(1)}-\epsilon_2)}}}L_k^+...L_2^+\psi^{(1)}_0(x),\\
&\psi_{k+n}^{(k)}(x)=\textstyle{\frac{1}{\sqrt{({\cal{E}}_{k+n-1}^{(k-1)}-\epsilon_k)...({\cal{E}}_n^--\epsilon_1)}}}L_k^+...L_1^+\psi^-_n(x). 
\end{split}
\end{equation}

\section{Graphene in particular magnetic fields}
In this section we are going to address two known shape invariant solvable potentials, which will be achieved by assuming that the magnetic field is either constant or has an exponential decay \cite{mrsr10,rr10,knn09,mf14}.

\subsection{Constant magnetic field}  

The constant magnetic field is obtained from a linear vector potential ${\bf A}=B_0x\hat{y}$, $B_0>0$. This case leads to the well known harmonic oscillator potentials, which in terms of the superpotential $W_0(x)=\frac{\omega}{2}x+k$ are given by (\ref{c2eq5}) with $\omega=2eB_0/c\hbar$. Now we choose $H^-$ as our departure Hamiltonian:

\begin{equation}
H^-=-\frac{d^2}{dx^2}+\frac{\omega^2}{4}\bigg(x+\frac{2k}{\omega}\bigg)^2-\frac{\omega}{2},
\end{equation}

\noindent
whose eigenvalues and eigenfunctions are given by:
\begin{equation}
{\cal{E}}_{n}^-=\omega n,~~~\psi^-_n(x)=N_ne^{-\frac{\omega}{4}(x+\frac{2k}{\omega})^2}H_n{\textstyle[\sqrt{\frac{\omega}{2}}(x+\frac{2k}{\omega})]},
\end{equation}

\noindent
where $n=0,1,2\dots$, the normalization factor is $N_n=\sqrt{\frac{1}{2^nn!}(\frac{\omega}{2\pi})^{\frac{1}{2}}}$ and $H_n$ are the Hermite polinomials. The first step of the method is to move up the energy of $H^-$ as follows:

\begin{equation}
\tilde{H}_0=-\frac{d^2}{dx^2}+\frac{\omega^2}{4}\bigg(x+\frac{2k}{\omega}\bigg)^2-\frac{\omega}{2}-\epsilon_1,
\end{equation}

\noindent
with $\epsilon_1\leq{\cal E}_0^-=0$. From the intertwining relation~(\ref{c3eq1}) it is obtained equation (\ref{c3eq2}). The change $W_1(x,\epsilon_1)=u_1^{(0)'}/u_1^{(0)}$ in the last equation leads to the Schr\"{o}dinger equation (\ref{c3eq10}), whose general solution is:  

\begin{equation}
u_1^{(0)}=e^{-\frac{\omega}{4}(x+\frac{2k}{\omega})^2}\bigg(\textstyle{{}_1\!F_1[a,\frac{1}{2},\frac{\omega}{2}(x+\frac{2k}{\omega})^2]+2\nu_1\frac{\Gamma[a+\frac{1}{2}]}{\Gamma[a]}\sqrt{\frac{\omega}{2}}(x+\frac{2k}{\omega}){}_1\!F_1[a+\frac{1}{2},\frac{3}{2},\frac{\omega}{2}(x+\frac{2k}{\omega})^2]}\bigg),
\label{c4eq1}
\end{equation}

\noindent
with $a=-\epsilon_1/2\omega$. Note that, in order to avoid singularities in the potential $V_1(x,\epsilon_1)$ the parameter $\nu_1$ must be restricted to the interval $(-1,1)$. In particular, for the parameters $\epsilon_1=-\omega/5$ and $\nu_1=0$ the superpotential becomes:

\begin{equation}
W_1(x,\epsilon_1)={\textstyle{\frac{\omega}{2}}(x+\frac{2k}{\omega})}\bigg(-1+\frac{2}{5}\frac{{}_1\!F_1[\frac{11}{10},\frac{3}{2},\frac{\omega}{2}(x+\frac{2k}{\omega})^2]}{{}_1\!F_1[\frac{1}{10},\frac{1}{2},\frac{\omega}{2}(x+\frac{2k}{\omega})^2]}\bigg),
\end{equation}

\noindent
and the new potential reads:

\begin{equation}
V_1(x,\epsilon_1)=\tilde{V}_0-2\frac{d}{dx}\bigg[{\textstyle\frac{\omega}{2}(x+\frac{2k}{\omega})}\bigg(-1+\frac{2}{5}\frac{{}_1\!F_1[\frac{11}{10},\frac{3}{2},\frac{\omega}{2}(x+\frac{2k}{\omega})^2]}{{}_1\!F_1[\frac{1}{10},\frac{1}{2},\frac{\omega}{2}(x+\frac{2k}{\omega})^2]}\bigg)\bigg].
\label{c4eq2}
\end{equation}

\noindent
In the variable $\zeta=\sqrt{\omega/2}(x+2k/\omega)$ it is obtained that:

\begin{equation}
V_1(\zeta,\epsilon_1)=\frac{\omega}{2}\zeta^2\bigg[1+\frac{8}{25}\bigg(\frac{{}_1\!F_1[\frac{11}{10},\frac{3}{2},\zeta^2]}{{}_1\!F_1[\frac{1}{10},\frac{1}{2},\zeta^2]}\bigg)^2-\frac{88}{75}\frac{{}_1\!F_1[\frac{21}{10},\frac{5}{2},\zeta^2]}{{}_1\!F_1[\frac{1}{10},\frac{1}{2},\zeta^2]}\bigg]-\frac{2}{5}\omega\frac{{}_1\!F_1[\frac{11}{10},\frac{3}{2},\zeta^2]}{{}_1\!F_1[\frac{1}{10},\frac{1}{2},\zeta^2]}+\frac{7}{10}\omega,
\end{equation}

\noindent
which in a more compact form reads as: 

\begin{equation}
V_1(\zeta,\epsilon_1)=\frac{\omega}{2}\zeta^2+\frac{3}{10}\omega-16\omega\zeta^2\bigg[\frac{{}_1\!F_1[\frac{1}{10},\frac{3}{2},\zeta^2]}{5({}_1\!F_1[\frac{1}{10},\frac{1}{2},\zeta^2])^2}\bigg({}_1\!F_1[\frac{1}{10},\frac{1}{2},\zeta^2]-\frac{4}{5}{}_1\!F_1[\frac{1}{10},\frac{3}{2},\zeta^2]\bigg)\bigg].
\end{equation}

\noindent
For this case the magnetic field is given by:

\begin{equation}
B_1(x,\epsilon_1)=-B_0+\frac{2B_0}{5}\frac{d}{dx}{\bigg[(x+\frac{2k}{\omega})}\frac{{}_1\!F_1[\frac{11}{10},\frac{3}{2},\frac{\omega}{2}(x+\frac{2k}{\omega})^2]}{{}_1\!F_1[\frac{1}{10},\frac{1}{2},\frac{\omega}{2}(x+\frac{2k}{\omega})^2]}\bigg].
\end{equation}

\begin{figure}[h]
\centering
\subfigure[]{\includegraphics[width=75mm]{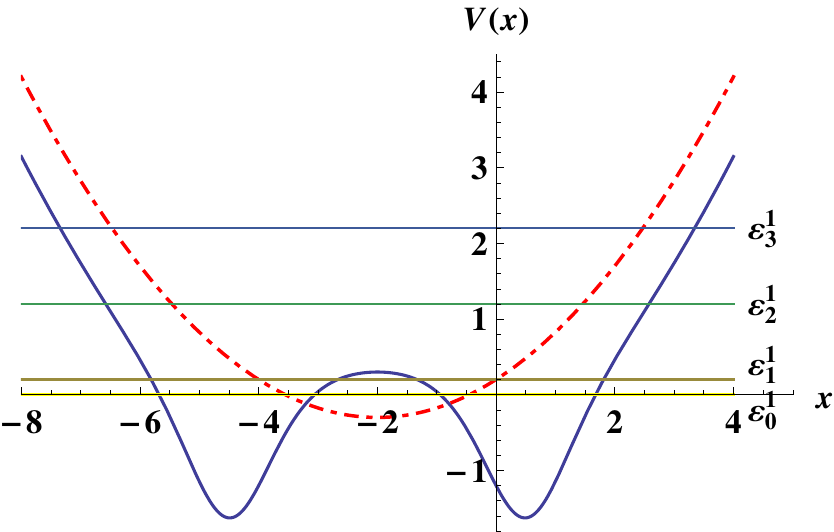}}
\subfigure[]{\includegraphics[width=75mm]{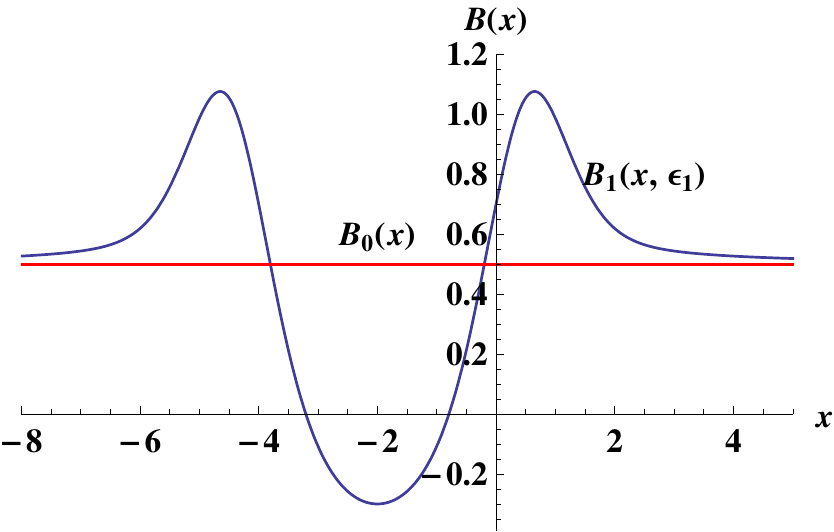}}
\caption{First intertwining for the harmonic oscillator: (a) generated potential $V_1(x,\epsilon_1)$ (continuous line) and the initial one $\tilde{V}_0$ (dotted line), whose energy levels are ${\cal{E}}^{(1)}_0=0$, ${\cal{E}}^{(1)}_1=\frac{1}{5}\omega$, ${\cal{E}}^{(1)}_2=\frac{6}{5}\omega$, ${\cal{E}}^{(1)}_3=\frac{11}{5}\omega$; (b) generated magnetic field $B_1(x,\epsilon_1)$, and the constant initial one $B_0$ (horizontal line) for $\omega=1$.} \label{fig2}
\end{figure}

\noindent
Some graphs of the potential $V_1(x,\epsilon_1)$ and associated magnetic field $B_1(x,\epsilon_1)$ are shown in figure~\ref{fig2}. Since the two lowest energy levels are very close, the form of $V_1(x,\epsilon_1)$ resembles the famous Mexican hat, which illustrates the spontaneous symmetry breaking in field theory. In our case the Mexican hat can be modified by changing $\epsilon_1$, and its existence means that the Dirac electron can pass from the ground to the first excited state with a greater probability than for any two other states.

\noindent
The eigenenergies of the system in our example are:
\begin{equation}
E_0^{(1)}=0,~~~E_{n+1}^{(1)}=\hbar v_F\sqrt{\omega\left(n+\frac{1}{5}\right)},~~~n=0,1,\dots
\end{equation}

\noindent
The corresponding eigenfunctions, taking into account (\ref{c2eq2}), are given by:

\begin{equation}
\begin{split}
&\Psi_0(x,y)= e^{iky}\left[\begin{array}{c}
0  \\ 
i\psi_0^{(1)}(x)
\end{array} \right]
\sim e^{iky}\left[\begin{array}{c}
0  \\ 
i\frac{1}{{}_1\!F_1[\frac{1}{10},\frac{1}{2},\frac{\omega}{2}(x+\frac{2k}{\omega})^2]}e^{\frac{\omega}{4}(x+\frac{2k}{\omega})^2} 
\end{array} \right],\\
&\Psi_{n+1}(x,y)= e^{iky}\left[\begin{array}{c}
\psi^-_n(x)  \\ 
i\psi^{(1)}_{n+1}(x)
\end{array} \right]
=e^{iky}\left[\begin{array}{c}
\psi^-_n(x) \\ 
i \frac{1}{\sqrt{\omega(n+1/5)}}L^+_1\psi^-_n(x) 
\end{array}\right].
\label{c4eq3}
\end{split}
\end{equation}

\begin{figure}[h]
\centering
\subfigure[]{\includegraphics[width=75mm]{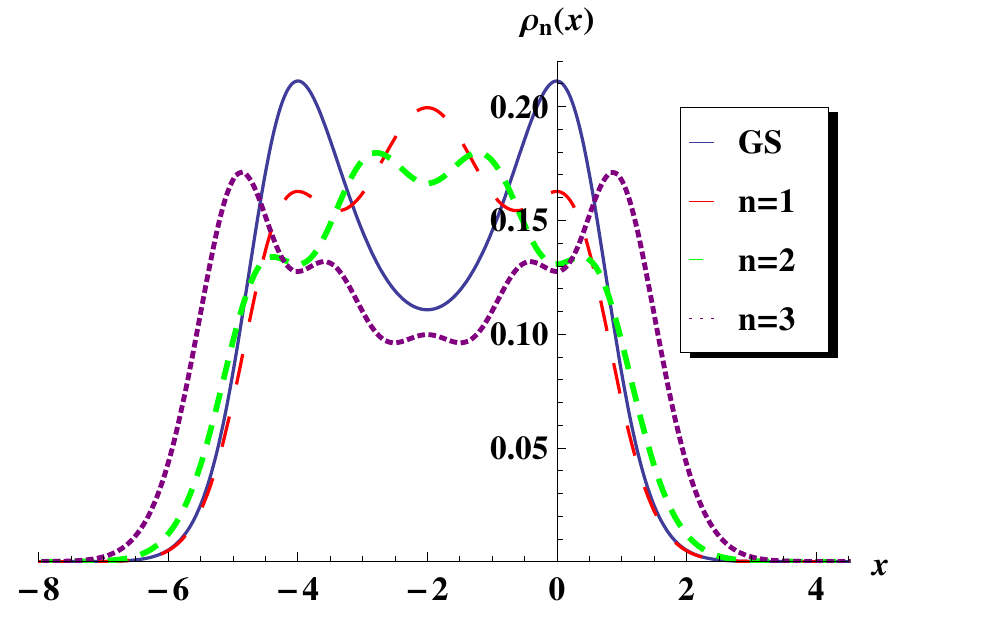}}
\subfigure[]{\includegraphics[width=75mm]{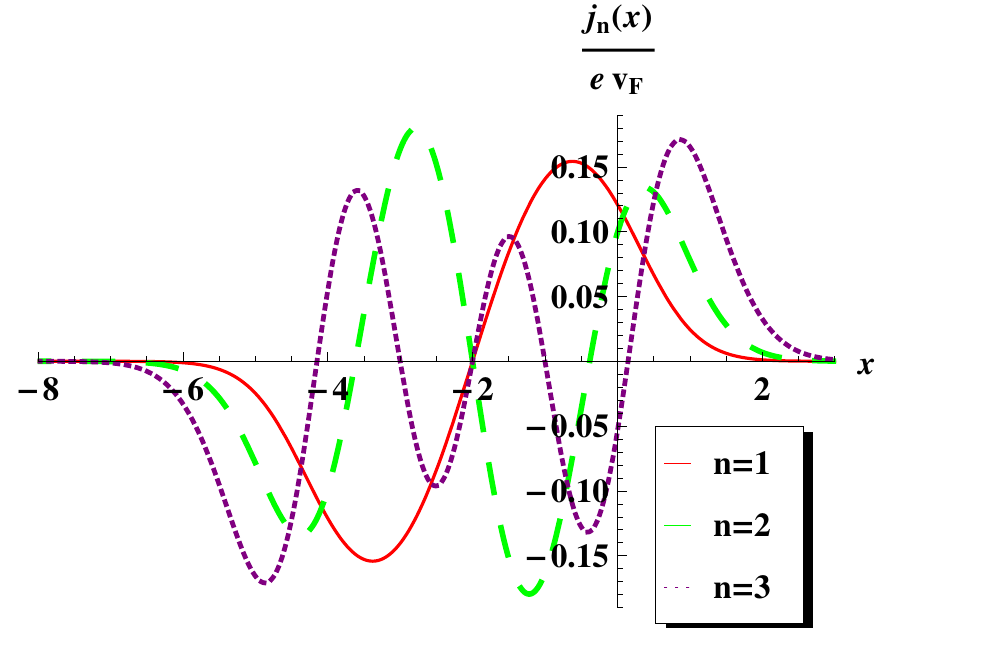}}
\caption{(a) Probability density $\rho_0(x)$ for the ground state (GS, blue) and the excited states $\rho_{n+1}(x)$ for $n=0,1,2$ (red, green, purple); (b) current densities for the correponding excited states for $n=0,1,2$ with the same colors that in (a), where $\omega=c=\hbar=e=1$.} \label{fig3}
\end{figure}

\noindent
With these expressions it is customary to calculate the probability density and probability current. The probability density for the excited states of graphene is $\rho_{n+1}(x)=|\psi_n^{(0)}(x)|^2+|\psi_{n+1}^{(1)}(x)|^2$ while for the ground state is $\rho_0(x)=|\psi_0^{(1)}(x)|^2$. The probability currents are $j_{n+1}(x)=ev_F\psi_n^{(0)}(x)\psi_{n+1}^{(1)}(x)$ and $j_0(x)=0$ respectively \cite{knn09}. Some graphs for the probability density and probability current are shown in figure~\ref{fig3}.

\noindent
It is seen that the greater probability appears around the points where a minimum of the potential exists. Moreover, since the potential is symmetric with respect to the point $-2k/\omega$, then the probability density has also the same symmetry.

\subsubsection{Second-order magnetic field}

In the second intertwining the potential~(\ref{c4eq2}) is moved up by an amount $-\epsilon_2$:

\begin{equation}
\tilde{V}_1(x,\epsilon_1)=\bigg[\frac{\omega^2}{4}\bigg(x+\frac{2k}{\omega}\bigg)^2-\frac{\omega}{2}-\epsilon_1-2\frac{d}{dx}W_1(x,\epsilon_1)\bigg]-\epsilon_2.
\end{equation}

\noindent
Then, it is built the second-order potential:

\begin{equation}
V_2(x,\epsilon_2)=\tilde{V}_0-2\frac{d^2}{dx^2}\ln \mathsf{W}[u_1^{(0)},u_2^{(0)}]-\epsilon_2.
\end{equation}

\noindent
Now it is required to find the general solution $u_2^{(0)}$ of Eq. (\ref{c3eq11}), which is the same of Eq. (\ref{c4eq1}) but the parameter $a$ is modified to $a_1=-(\epsilon_1+\epsilon_2)/2\omega$. Then we find $u_2^{(1)}\propto L_1^+u_2^{(0)}$ and we build the operators for the second intertwining, $L_2^\pm=\mp d/dx+u_2^{(1)'}/u_2^{(1)}$. For this example we choose the particular values $\epsilon_1=-\omega/5$ and $\epsilon_2=-3\omega$. Therefore, the eigenenergies of the system become:

\begin{equation}
E_0^{(2)}=0,~~~E_{n+1}^{(2)}=\hbar v_F\sqrt{{\cal E}_n^{(1)}+3\omega},~~~n=0,1,\dots
\end{equation}

\noindent
The corresponding two-component spinors are:

\begin{equation}
\begin{split}
\Psi_0(x,y)=e^{iky}\left[\begin{array}{c}
0  \\ 
i\psi_0^{(2)}(x)
\end{array} \right],~~~ 
\Psi_1(x,y)=e^{iky}\left[\begin{array}{c}
\psi_0^{(1)}(x)\\ 
i\psi_1^{(2)}(x)
\end{array} \right],~~~
\Psi_{n+2}(x,y)=e^{iky}\left[\begin{array}{c}
\psi^{(1)}_{n+1}(x)\\ 
i \psi^{(2)}_{n+2}(x)
\end{array} \right],
\end{split}
\end{equation}

\noindent
where

\begin{equation}
\begin{split}
\psi^{(2)}_{n+2}(x)=\frac{L^+_2\psi^{(1)}_{n+1}(x)}{\sqrt{({\cal{E}}_{n+1}^{(1)}-\epsilon_2)}},~~~\psi_1^{(2)}(x)=\frac{L^+_2\psi_0^{(1)}(x)}{\sqrt{({\cal{E}}_0^{(1)}-\epsilon_2)}},~~~\psi_0^{(2)}(x)\sim \frac{1}{u^{(1)}_2}.
\end{split}
\end{equation}

\begin{figure}[h]
\centering
\subfigure[]{\includegraphics[width=75mm]{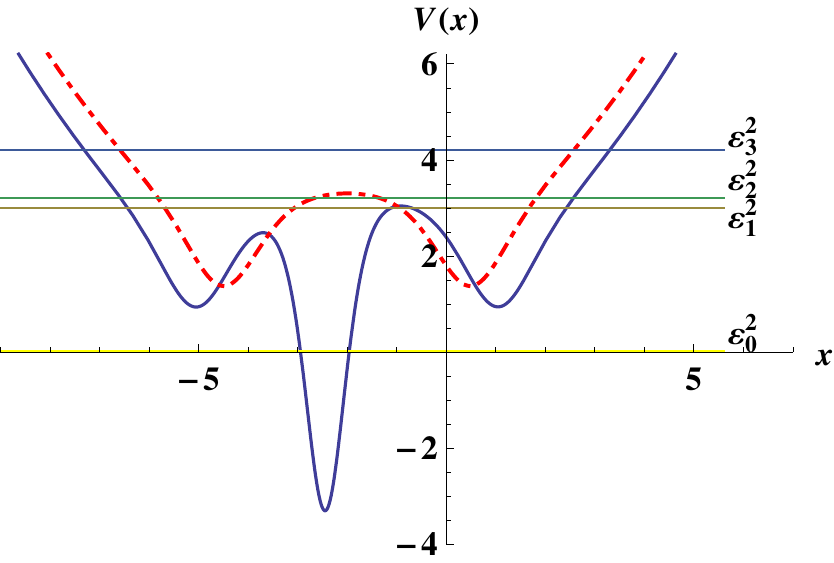}}
\subfigure[]{\includegraphics[width=75mm]{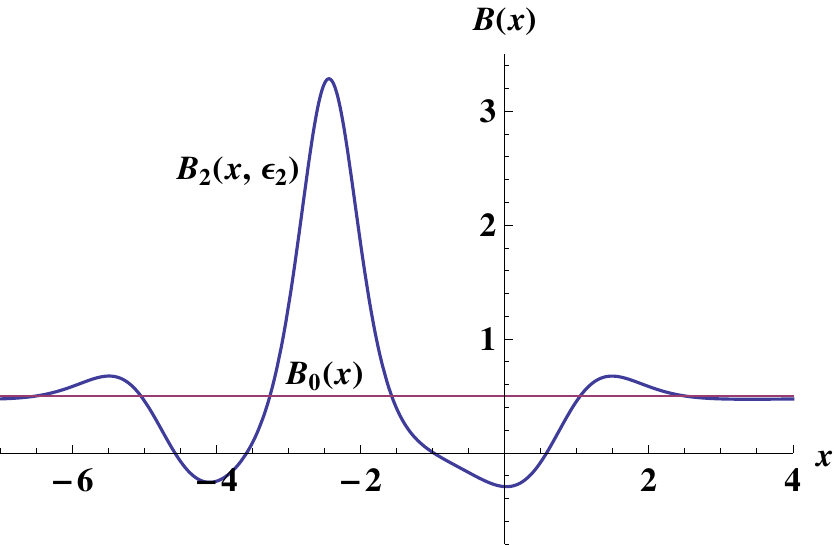}}
\subfigure[]{\includegraphics[width=75mm]{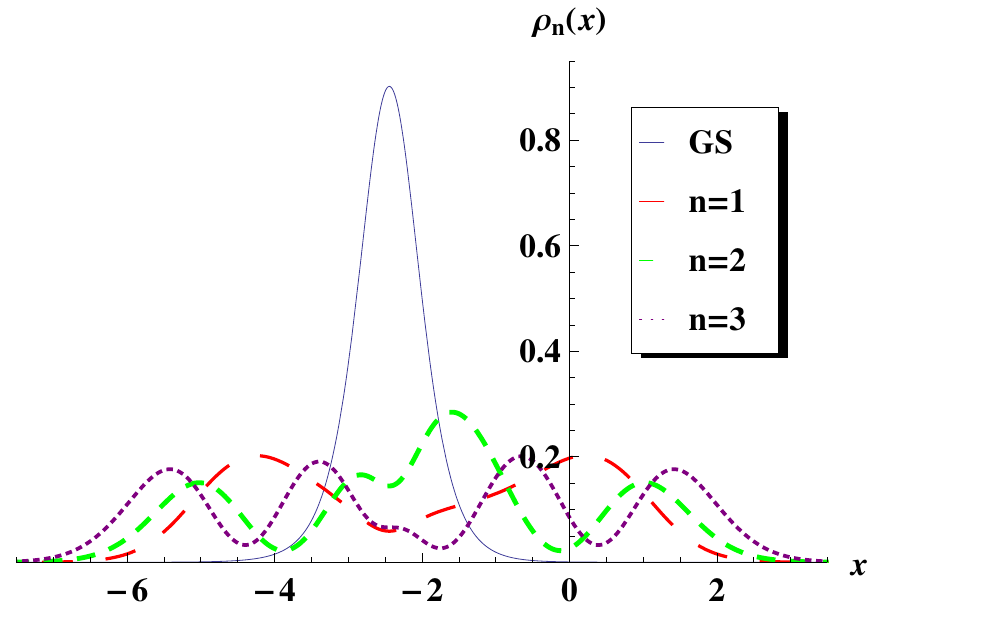}}
\subfigure[]{\includegraphics[width=75mm]{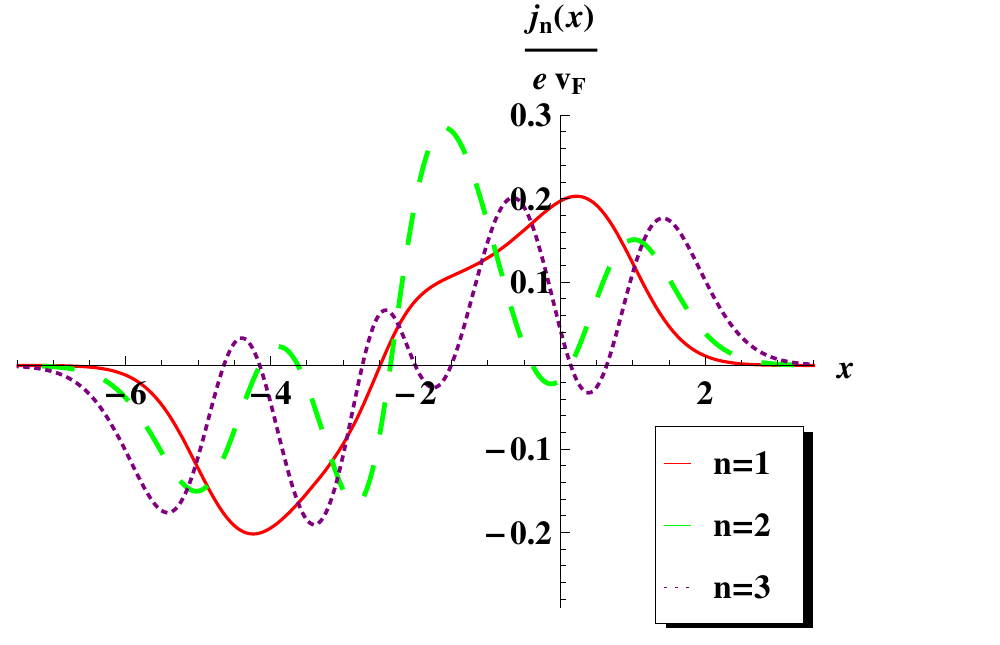}}
\caption{Second intertwining for the harmonic oscillator with $\nu_1=0$, $\nu_2=\frac{3}{2}$, $\epsilon_1=-\frac{1}{5}\omega$, $\epsilon_2=-3\omega$, $\omega=1$: (a) generated potential $V_2(x,\epsilon_2)$ (continuous line) and initial one $\tilde{V_1}(x,\epsilon_1)$ (dashed line), with the energy levels ${\cal{E}}^{(2)}_0=0$, ${\cal{E}}^{(2)}_1=3\omega$, ${\cal{E}}^{(2)}_2=\frac{16}{5}\omega$, ${\cal{E}}^{(2)}_3=\frac{21}{5}\omega$; (b) generated magnetic field $B_2(x,\epsilon_2)$ (blue), and uniform one (horizontal line); (c) probability densities for the ground state (GS, blue) and the excited states $|\psi_{n+1}^{(2)}(x)|^2$ for $n=0,1,2$ (red, green, purple); (d) current densities for the excited states $n=0,1,2$ with the same colors that in (c).} 
\label{fig3a}
\end{figure}

\noindent
Some plots for this example are shown in figure \ref{fig3a}, where it is drawn the second-order potential, magnetic field, probability density and current density. As in the first transformation, the shapes of the potential and magnetic field look closer to the standard parabola and horizontal line at the edges of the graphs. 

\noindent
Let us note that, in order to avoid singularities in $V_2$, the parameter $\nu_2$ must fullfill $\nu_2\in\mathbb{R}-\{(-1,1)\}$, which is the opposite to the restriction for $\nu_1$ of the first intertwining. This choice for $\nu_2$ guarantees that none singularity will appear in the new potential $V_2$.

\subsection{Exponentially decaying magnetic field}

In this example the vector potential reads ${\bf A}(x)=-(B_0/\alpha)e^{-\alpha x}\hat{y}$, with $B_0>0$, $\alpha>1$, thus the associated magnetic field becomes $B_0(x)=B_0e^{-\alpha x}$. The superpotential in this case is given by $W_0=k-De^{-\alpha x}$, $D=eB_0/c\hbar\alpha$, which leads to the Morse potentials:

\begin{equation}
V^{\pm}(x)=k^2+D^2e^{-2\alpha x}-2D\bigg(k\mp\frac{\alpha}{2}\bigg)e^{-\alpha x}.
\end{equation}

\noindent
We choose $V^-(x)$ and displace it by $-\epsilon_1$ to produce $\tilde{V}_0(x)$; the new potential $V_1(x,\epsilon_1)$ depends on $W_1(x,\epsilon_1)$, which is a solution of the Riccati Eq. (\ref{c3eq2}). The new superpotential is written as $W_1(x,\epsilon_1)=u_1^{(0)'}/u_1^{(0)}$, with $u_1^{(0)}$ being the general solution of the Schr\"{o}dinger equation (\ref{c3eq10}) given by:

\begin{equation}
u_1^{(0)}=e^{-\frac{D}{\alpha}e^{-\alpha x}}e^{-\sqrt{k^2-\epsilon_1}x}\bigg({}_1F_1[a,b,\textstyle{\frac{2D}{\alpha}e^{-\alpha x}}]+(\frac{2k}{\alpha})(1+\frac{1}{\nu_1})U[a,b,\textstyle{\frac{2D}{\alpha}e^{-\alpha x}}]\bigg),
\end{equation}

\noindent
where $\nu_1$ obeys the restriction $\nu_1\in\mathbb{R}-\{[-1,0]\}$ and the parameters $a$, $b$ are defined as:

\begin{equation}
a=-\frac{k}{\alpha}+\frac{\sqrt{k^2-\epsilon_1}}{\alpha},~~~b=1+2\frac{\sqrt{k^2-\epsilon_1}}{\alpha}.
\end{equation}

\noindent
Therefore, the superpotential turns out to be:

\begin{equation}
W_1(x,\epsilon_1)=De^{-\alpha x}-\sqrt{k^2-\epsilon_1}+{\cal F}(x),
\end{equation}

\noindent    
where the function ${\cal F}(x)$ reads:

\begin{equation} 
{\cal F}(x)=-\left(\frac{2aDe^{-\alpha x}}{b}\right)\frac{{}_1F_1[1+a,1+b,\frac{2D}{\alpha}e^{-\alpha x}]-b(\frac{2k}{\alpha})(1+\frac{1}{\nu_1})U[1+a,1+b,\frac{2D}{\alpha}e^{-\alpha x}]}{{}_1F_1[a,b,\frac{2D}{\alpha}e^{-\alpha x}]+(\frac{2k}{\alpha})(1+\frac{1}{\nu_1})U[a,b,\frac{2D}{\alpha}e^{-\alpha x}]}.
\end{equation}

\noindent
The new potential and associated magnetic field are given by:

\begin{equation}
V_1(x,\epsilon_1)=\tilde{V}_0(x)-2\frac{d}{dx}[{\cal F}(x)+De^{-\alpha x}],~~~
B_1(x,\epsilon_1)=-B_0e^{-\alpha x}+\frac{c\hbar}{e}\frac{d}{dx}{\cal F}(x).
\end{equation}

\begin{figure}[h]
\centering
\subfigure[]{\includegraphics[width=75mm]{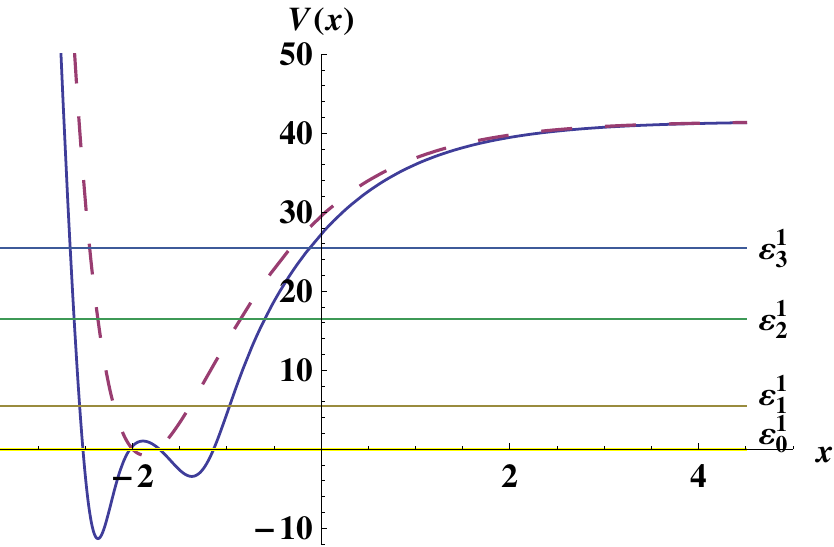}}
\subfigure[]{\includegraphics[width=75mm]{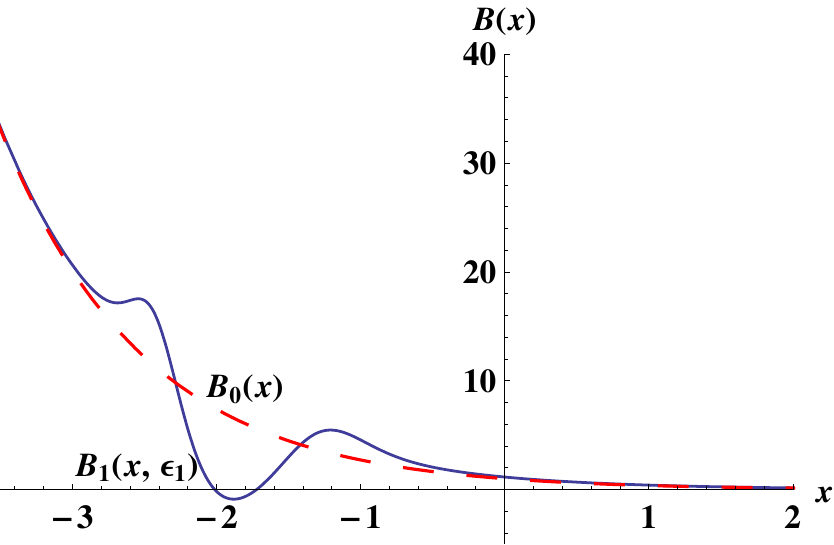}}
\caption{First intertwining for the Morse potential with $\nu_1=-\frac{3}{2}$, $k=6\alpha$, $\epsilon_1=-\frac{1}{2}{\cal{E}}_1^-=-\frac{11\alpha^2}{2}$, $\alpha=1$. (a) New potential $V_1(x,\epsilon_1)$ (continuous line) and initial one $\tilde{V_0}(x)$ (dashed line), with the energy levels at ${\cal{E}}^{(1)}_0=0$, ${\cal{E}}^{(1)}_1=\frac{11}{2}\alpha^2$, ${\cal{E}}^{(1)}_2=\frac{33}{2}\alpha^2$, ${\cal{E}}^{(1)}_3=\frac{51}{2}\alpha^2$; (b) generated magnetic field $B_1(x,\epsilon_1)$, and the initially decaying magnetic field (dashed line).} 
\label{fig4}
\end{figure}

\noindent
To illustrate these expressions we choose the factorization energy as $\epsilon_1=-{\cal{E}}_1^-/2=-\alpha(2k-\alpha)/2$: the corresponding plots can be observed in figure \ref{fig4}.

\noindent
The eigenenergies for the problem are given by:

\begin{equation}
{\cal{E}}_0^{(1)}=0,~~~{\cal{E}}_{n+1}^{(1)}=\tilde{\cal{E}}_n^{(0)}=\alpha n(2k-\alpha n)+\frac{\alpha}{2}(2k-\alpha),~~~n=0,1,\dots
\end{equation} 

\noindent
The eigenfunctions corresponding to $H_1$ take the form:

\begin{equation}
\begin{split}
&\psi_0^{(1)}(x)\sim \frac{e^{\frac{D}{\alpha}e^{-\alpha x}}e^{\sqrt{k^2-\epsilon_1}x}}{{}_1F_1[a,b,\textstyle{\frac{2D}{\alpha}e^{-\alpha x}}]+(\frac{2k}{\alpha})(1+\frac{1}{\nu_1})U[a,b,\textstyle{\frac{2D}{\alpha}e^{-\alpha x}}]},\\
&\psi^{(1)}_{n+1}(x)=\frac{1}{\sqrt{\alpha[n(2k-\alpha n)+\frac{1}{2}(2k-\alpha)]}}L^+_1(x,\epsilon_1)\psi^-_n(x),~~~n=0,1,\dots
\end{split}
\end{equation}

\noindent
where $\psi^-_n(x)$ is an eigenfunction of $\tilde{H}_0$ given by:
\begin{equation}
\psi_n^-(x)=N_ne^{-\frac{D}{\alpha}e^{-\alpha x}-(k-\alpha n)x}L_n^{2(k/\alpha)-2n}(\textstyle{\frac{2D}{\alpha}e^{-\alpha x}}).
\end{equation}

\noindent
In this equation the normalization constant reads $N_n=(2D/\alpha)^{(k/\alpha-n)}\sqrt{2\alpha(k/\alpha- n)n!/(2k/\alpha- n)!}$, while $L_n^{(\alpha)}(x)$ are the Laguerre polynomials, $n=0,1,2\dots\leq k$. The spinor expressions are similar to those of Eqs. (\ref{c4eq3}).\\

\begin{figure}[h]
\centering
\subfigure[]{\includegraphics[width=75mm]{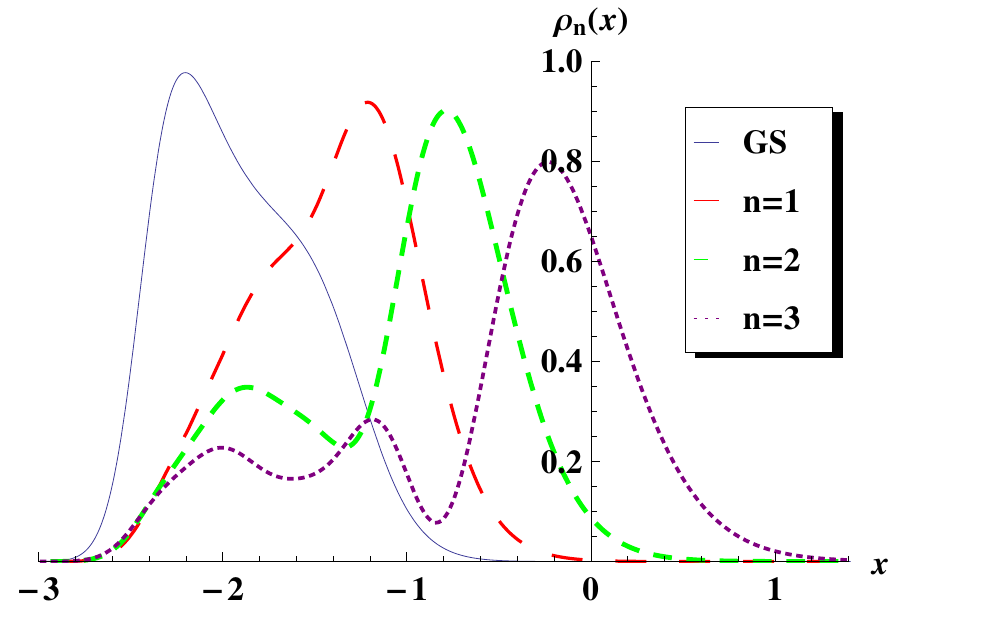}}
\subfigure[]{\includegraphics[width=75mm]{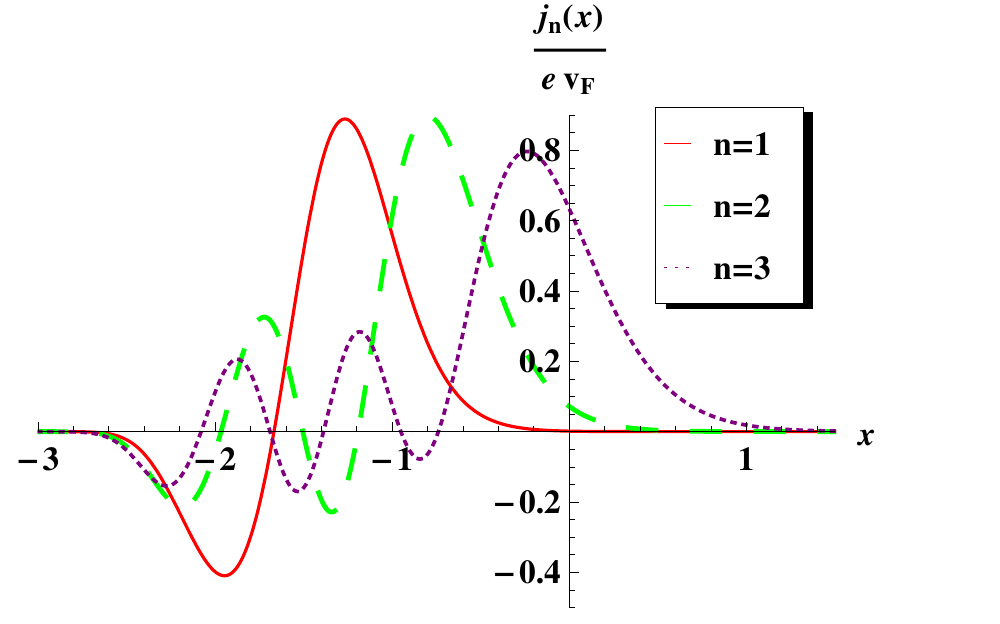}}
\caption{(a) Probability density $|\Psi_{n}^{(1)}(x)|^2$ for the ground state (GS, blue) and for $n=1,2,3$ (red, green, purple); (b) current density for the excited states $n=1,2,3$ with the same colors that in (a). The parameters taken are the same that in figure 5.} 
\label{fig4a}
\end{figure}

\noindent
Some plots for the probability and current density are shown in figure \ref{fig4a}. 

\subsubsection{Second-order exponentially decaying magnetic field}

\begin{figure}[h]
\centering
\subfigure[]{\includegraphics[width=75mm]{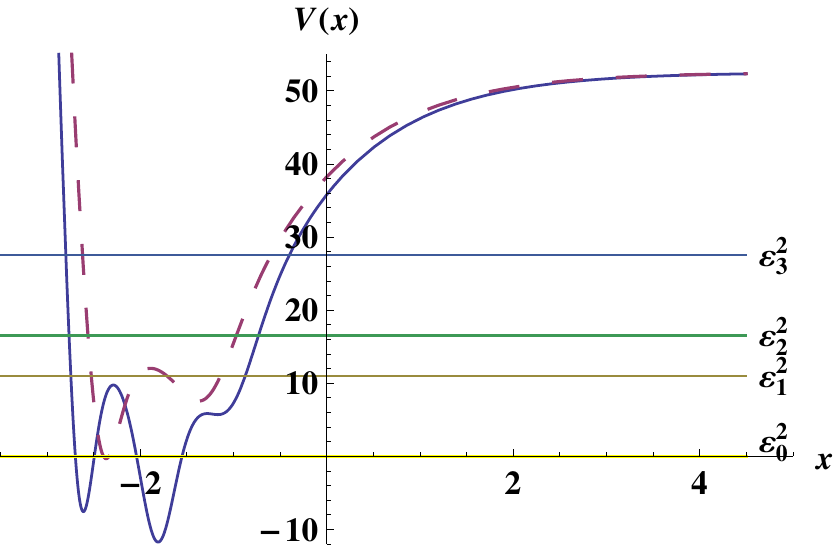}}
\subfigure[]{\includegraphics[width=75mm]{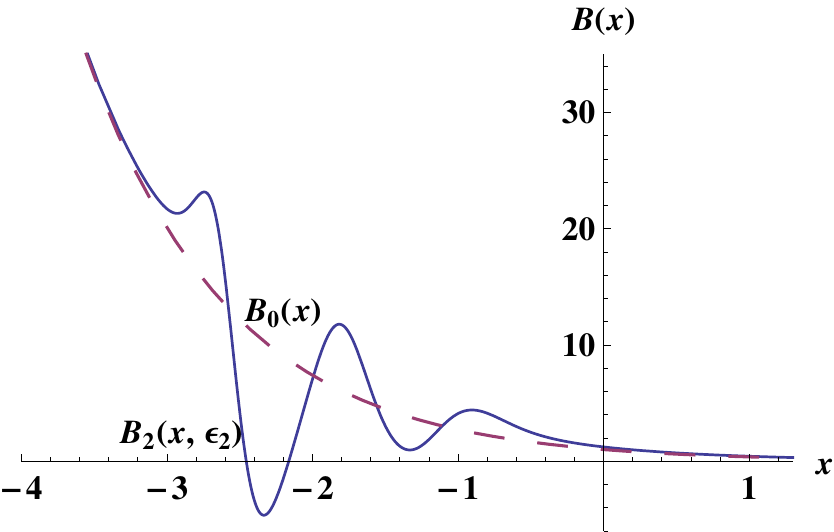}}
\subfigure[]{\includegraphics[width=75mm]{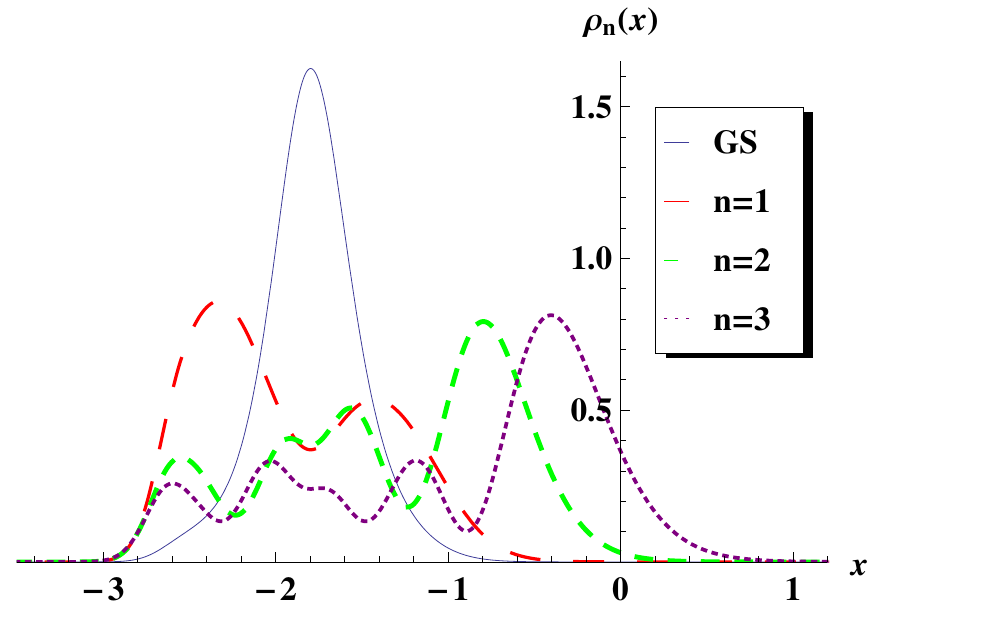}}
\subfigure[]{\includegraphics[width=75mm]{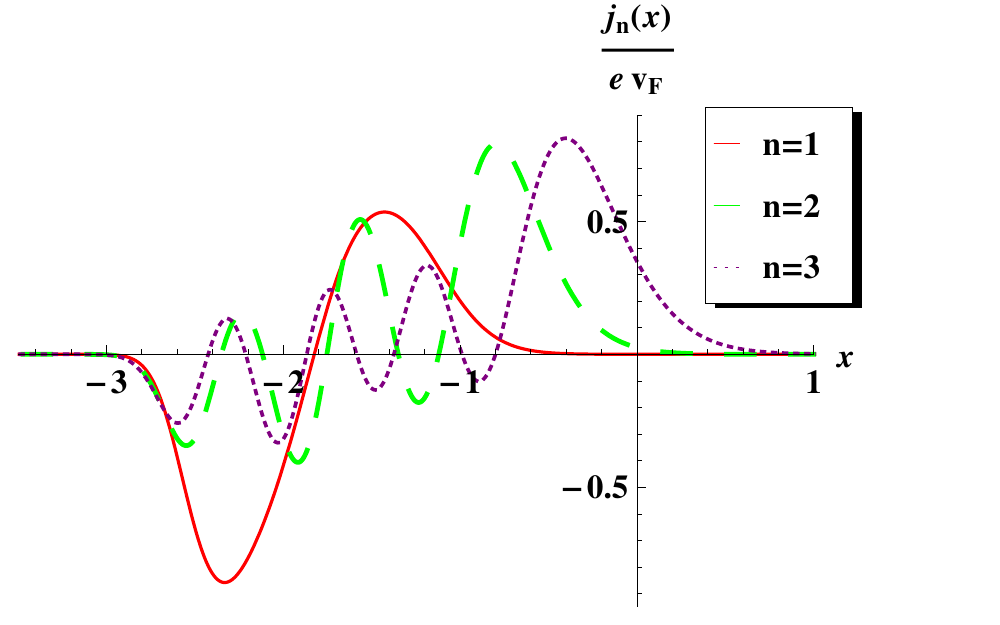}}
\caption{Second intertwining for the Morse potential with $k=6\alpha$, $\nu_1=-\frac{3}{2}$, $\nu_2=-\frac{1}{2}$, $\epsilon_1=-\frac{1}{2}{\cal{E}}_1^-=-\frac{11\alpha^2}{2}$, $\epsilon_2=-{\cal{E}}_1^-=-11\alpha^2$: (a) generated potential $V_2(x,\epsilon_2)$ (continuous line) and initial one $\tilde{V_1}(x,\epsilon_1)$ (dashed line), with energy levels ${\cal{E}}^{(2)}_0=0$, ${\cal{E}}^{(2)}_1=11\alpha^2$, ${\cal{E}}^{(2)}_2=\frac{33\alpha^2}{2}$, ${\cal{E}}^{(2)}_3=\frac{55\alpha^2}{2}$; (b) generated magnetic field $B_2(x,\epsilon_2)$, and initially decaying magnetic field (dashed line); (c) probability density $|\Psi_{n}^{(2)}(x)|^2$ for the ground state (GS, blue) and the excited states $n=1,2,3$ (red, green, purple); (d) current density for the excited states with the same colors that in (c).} 
\label{fig5}
\end{figure}

This case is similar to the second intertwining for the harmonic oscillator, namely, the potential $V_1(x,\epsilon_1)$ is moved up a quantity $-\epsilon_2={\cal{E}}_1^-$, and we choose $-\epsilon_1={\cal{E}}_1^-/2$, where ${\cal{E}}_1^-$ is the first excited state for the Morse potential $V^-(x)$. The transformation parameters are now $\nu_1=-3/2$, and $\nu_2=-1/2$. Note that the appropriate values for $\nu_2$ are now $\nu_2\in[-1,0]$, which is the opposite to the restriction for $\nu_1$ found in the first intertwining. The corresponding graphs for the potential, magnetic field, probability density and current density are displayed in figure \ref{fig5}.

\section{Concluding remarks}\label{conclusion}

The first-order intertwining method presented here generalizes the shape invariant technique that different authors have used previously to describe the behavior of an electron near to the Dirac points in a graphene layer with applied external magnetic fields \cite{knn09,mtm11,chrv18,jss18,jk14,j15}. Similarly as in \cite{mf14}, we have modified the form of the magnetic field without destroying the exact solvability of the system, by using Schr\"{o}dinger seed solutions instead of solutions for the associated Riccati equation. Once again, we recover the shape invariant results, but now in the limit $\epsilon_1\rightarrow0$. We have shown as well that the factorization energy $\epsilon_1$ and the parameter $\nu_1$ control the deformations induced in the initial form of the potential, magnetic field, probability density and current density.\\

In addition, we have introduced an iterative procedure for generating new families of higher-order potentials and magnetic fields, thus we know how to find the $k$th-order potentials from the initial shape invariat one. The examples addressed through the second intertwining show more abrupt deformations with respect to the first transformation, but it is possible to control such deformations by appropriately choosing the four parameters $\epsilon_{1}$, $\epsilon_{2}$ and $\nu_{1}$, $\nu_{2}$.\\

We think that it is possible in the future to develop specific models making use of the methods described in this article, for example, due to the local confinement that can be achieved through the auxiliary potentials, they could be important in electronics research. Finally, it is hoped that this work would become a guide for colleagues interested in theoretical developments of graphene.


\section*{Acknowledgments}
The authors acknowledge the support of Conacyt, grant 284489. Miguel Castillo-Celeita also acknowledges the Conacyt fellowship 301117.


\end{document}